\documentclass[conference]{IEEEtran}
\ifCLASSINFOpdf
\else
\fi
\usepackage{hyperref}
\usepackage{amsmath}
\usepackage{array}
\usepackage{graphicx} 
\usepackage{newtxtext,newtxmath}

\begin{document}
%
\title{A Comparative Evaluation of Deep Learning Models for Speech Enhancement in Real-World Noisy Environments}

\author{
\IEEEauthorblockN{Md Jahangir Alam Khondkar}
\IEEEauthorblockA{\small ECE, Clarkson University\\
Potsdam, NY, USA\\
khondkma@clarkson.edu}
\and
\IEEEauthorblockN{Ajan Ahmed}
\IEEEauthorblockA{\small ECE, Clarkson University\\
Potsdam, NY, USA\\
aahmed@clarkson.edu}

\and
\IEEEauthorblockN{Stephanie Schuckers}
\IEEEauthorblockA{\small ECE, Clarkson University\\
Potsdam, NY, USA\\
sschucke@clarkson.edu}
\and
\IEEEauthorblockN{Masudul Haider Imtiaz}
\IEEEauthorblockA{\small ECE, Clarkson University\\
Potsdam, NY, USA\\
mimtiaz@clarkson.edu}
}


%


\maketitle

\begin{abstract}
Speech enhancement, particularly denoising, is vital in improving the intelligibility and quality of speech signals for real-world applications, especially in noisy environments. While prior research has introduced various deep learning models for this purpose, many struggle to balance noise suppression, perceptual quality, and speaker-specific feature preservation, leaving a critical research gap in their comparative performance evaluation. This study benchmarks three state-of-the-art models—Wave-U-Net, CMGAN, and U-Net—on diverse datasets such as SpEAR, VPQAD, and Clarkson datasets. These models were chosen due to their relevance in the literature and code accessibility. The evaluation reveals that U-Net achieves high noise suppression with SNR improvements of +71.96\% on SpEAR, +64.83\% on VPQAD, and +364.2\% on the Clarkson dataset. CMGAN outperforms in perceptual quality, attaining the highest PESQ scores of 4.04 on SpEAR and 1.46 on VPQAD, making it well-suited for applications prioritizing natural and intelligible speech. Wave-U-Net balances these attributes with improvements in speaker-specific feature retention, evidenced by VeriSpeak score gains of +10.84\% on SpEAR and +27.38\% on VPQAD. This research indicates how advanced methods can optimize trade-offs between noise suppression, perceptual quality, and speaker recognition. The findings may contribute to advancing voice biometrics, forensic audio analysis, telecommunication, and speaker verification in challenging acoustic conditions.
\end{abstract}


%
\IEEEpeerreviewmaketitle

\section{Introduction}
Speech enhancement is essential for improving the intelligibility and quality of speech signals, particularly in noisy environments such as telephony, forensic audio analysis, and voice biometrics. In real-world applications, background noise and environmental distortions can significantly impact speech clarity, making it difficult for both humans and automated systems to process audio effectively. By applying advanced deep learning techniques, speech enhancement methods aim to suppress noise while preserving essential speech characteristics \cite{Maher2018}.

Recent advances in computational methods, such as deep learning (DL), have transformed the field of audio enhancement by providing models capable of learning complex representations of noise and speech. Using architectures such as encoder-decoders and attention mechanisms, these models have demonstrated substantial improvements in removing noise and improving speech quality. Unlike traditional methods, deep learning approaches generally adapt dynamically to various noise conditions \cite{Michelsanti2021}.

This study evaluates three such state-of-the-art DL models for speech enhancement, Wave-U-Net \cite{Stoller2018}, CMGAN (Convolutional-Multihead Attention Generative Adversarial Network) \cite{Abdulatif2024}, and Hybrid U-Net \cite{Belz2024}, for their unique capabilities of denoising and enhancing audio. Wave-U-Net, originally developed for music source separation, has proven effective in speech enhancement by leveraging an encoder-decoder architecture with skip connections \cite{Stoller2018}. This structure enables the model to learn both high-level abstractions and detailed temporal features, making it suitable for separating noise from speech in scenarios where preserving temporal details is crucial.
CMGAN introduces convolutional layers and transformer-based attention mechanisms. These components allow the model to dynamically focus on the most relevant parts of the audio signal while suppressing noise \cite{Abdulatif2024}. By incorporating a discriminator through its GAN framework, CMGAN ensures that the enhanced audio maintains a natural and realistic quality, which is particularly beneficial for handling non-stationary noise conditions \cite{Zhang2022}.

The U-Net model combines traditional convolutional components with advanced feature extraction techniques, capturing intricate details of noisy speech \cite{Belz2024}. This model enhances speech quality even in challenging noise environments by integrating multiple feature extraction stages and using skip connections. Historically, these models have evolved through various innovations, such as attention mechanisms, dilated convolutions, and probabilistic enhancements, enabling their application in diverse and complex real-world audio scenarios. These advancements have made them powerful tools for denoising audio in speaker recognition and forensic applications.

Existing speech enhancement models face significant challenges in handling non-stationary and diverse real-world noise conditions (including environmental noise) and generalization limitations. The model evaluations mostly rely on datasets used during training, which can introduce bias and fail to reflect true model performance in unseen conditions. Additionally, the lack of systematic testing incorporating commercial-grade speaker recognition metrics limits understanding their effectiveness in applications such as forensic analysis and speaker verification. \textbf{This study addresses these gaps through the following contributions:}

a) We systematically compare three state-of-the-art deep learning models for speech enhancement—Wave-U-Net, CMGAN, and U-Net—under standardized evaluation conditions. This ensures the models are fairly compared using identical evaluation parameters and datasets.

b) We establish a benchmark by utilizing diverse datasets, including SpEAR, VPQAD, and Clarkson, to assess how well these models perform in real-world noisy environments. By incorporating various noise conditions, we ensure that the findings are robust and applicable to practical scenarios.

c) We integrate speaker recognition metrics, such as VeriSpeak, alongside traditional speech enhancement metrics like PESQ and SNR. This allows us to examine the impact of speech enhancement on biometric applications and ensure that the enhanced audio maintains its usability for speaker recognition tasks.

Finally, we provide insights into the trade-offs between noise suppression, perceptual quality, and speaker-specific feature preservation across the different models. This analysis serves as a valuable resource for researchers and practitioners seeking to develop improved speech enhancement techniques that effectively balance these critical aspects.

The rest of the paper is organized as follows: Section 2 describes the datasets used for training and evaluation. Section 3 outlines the model architectures and training methodologies. Section 4 presents the results, while Section 5 discusses the findings, focusing on strengths, limitations, and future directions. This comparative analysis will help understand the strengths and areas for future research across these approaches.

 

\subsection{Related Work}
Table 1 provides a comparative analysis of the related work, including each model's performance and involved datasets. One of the recent and efficient DL models in the current domain is Wave-U-Net, proposed by Stoller et al. \cite{Stoller2018}, which uses an encoder-decoder structure with skip connections to separate speech from noise effectively. Subsequent works has built on the success of Wave-U-Net by incorporating attention mechanisms and adversarial training. Macartney and Weyde \cite{Macartney2018} applied the Wave-U-Net architecture for speech enhancement, showing improvements in standard metrics such as PESQ (Perceptual Evaluation of Speech Quality) and SSNR (Segmental Signal-to-Noise Ratio) with the training dataset VCTK (Voice Cloning Toolkit) and DEMAND (Diverse Environments Multichannel Acoustic Noise Database). Ali et al. \cite{Ali2020} introduced a dilated version of Wave-U-Net, with the VCTK and Librispeech training data set utilizing dilated convolutions to capture a broader temporal context without significantly increasing computational complexity. An attention-based version of Wave-U-Net was also developed to enhance speech quality by dynamically focusing on the most important audio signal features by Giri et al. \cite{Giri2019}.

\begin{table*}[htbp]
\centering
\small
\caption{Summary of Related Work in Speech Enhancement Using Wave-U-Net, CMGAN, and U-Net Models. Our work uniquely includes systematic model comparison, incorporates independent real-world evaluation datasets (SpEAR, VPQAD, Clarkson), and evaluates speaker recognition impacts (VeriSpeak), distinguishing it from prior studies.}
\label{tab:related_work}
\renewcommand{\arraystretch}{1.5} 
\setlength{\tabcolsep}{1.5pt} 

\begin{tabular}{>{\centering\arraybackslash}p{2.8cm}|
                >{\centering\arraybackslash}p{3.2cm}|
                >{\centering\arraybackslash}p{3.5cm}|
                >{\centering\arraybackslash}p{3.5cm}|
                >{\centering\arraybackslash}p{3.8cm}}

\hline
\textbf{Authors}         & \textbf{Deep Learning Model Used} & \textbf{Training Dataset} & \textbf{Evaluation Dataset} & \textbf{Evaluation Metrics} \\ \hline
Stoller et al. \cite{Stoller2018}       & Wave-U-Net, A multiscale NN       & MUSDB18, CCMixter           & MUSDB18                       & SDR                         \\ \hline
Macartney et al. \cite{Macartney2018}     & Wave-U-Net                       & VCTK                      & VCTK with DEMAND            & PESQ, CSIG, CBAK, COVL, SSNR \\ \hline
Ali et al. \cite{Ali2020}           & Dilated Wave-U-Net               & VCTK, Librispeech         & VCTK, Librispeech           & PESQ, STOI, SNR             \\ \hline
Giri et al. \cite{Giri2019}          & Attention Wave-U-Net             & VCTK                      & VCTK                        & PESQ, CSIG, CBAK, COVL, SSNR \\ \hline
Abdullah et al. \cite{Abdulatif2024}      & CMGAN                            & VCTK                      & VCTK with DEMAND            & PESQ, CSIG, CBAK, COVL, STOI, SSNR \\ \hline
Dai et al. \cite{Dai2023}         & CMGAN                            & VoiceBank-DEMAND          & VoiceBank-DEMAND            & PESQ, STOI                  \\ \hline
Zhang et al. \cite{Zhang2024}        & CMGAN                            & VCTK, Urban Sound         & VCTK, Urban Sound           & PESQ, CSIG, CBAK, COVL, SSNR, STOI \\ \hline
Belz et al. \cite{Belz2024}          & U-Net                            & LibriSpeech, ESC-50       & LibriSpeech, ESC-50         & Not Mentioned               \\ \hline
Baloch et al. \cite{Baloch2023}       & U-Net                            & Valentini, DEMAND         & Valentini, DEMAND           & PESQ, STOI                  \\ \hline
Nustede et al. \cite{Nustede2021}      & Variational U-Net                & MS-SNSD, DEMAND           & MS-SNSD                     & PESQ, STOI                  \\ \hline
Hossain et al. \cite{UNET2024}      & U-Net                            & IEEE Corpus, NOISEX-92    & IEEE Corpus                 & HASQI, PESQ, STOI           \\ \hline
\textbf{Our Work}        & \textbf{Wave-U-Net, CMGAN, U-Net} & \textbf{Wave-U-Net (MUSDB18 HQ + DEMAND [chosen for music/speech separation]), CMGAN (VCTK + DEMAND [chosen for multi-speaker clarity]), U-Net (LibriSpeech + ESC-50 + DEMAND [chosen for diverse speech/noise scenarios])} & \textbf{SpEAR [publicly available artificial noisy dataset], VPQAD [real-world noisy adults, 18-40 years], Clarkson Dataset [real-world noisy children, 4-18 years; unique for age and environment diversity]} & \textbf{PESQ, SNR, VeriSpeak (speaker recognition)}        \\ \hline
\end{tabular}

\end{table*}

The Convolutional-Multihead Attention Generative Adversarial Network (CMGAN), introduced by Abdullatif et al. \cite{Abdulatif2024}. CMGAN combines convolutional layers with transformer-based multi-head attention to dynamically focus on relevant parts of the speech signal while suppressing noise. The addition of GANs helped improve the naturalness of the output, making the enhanced speech sound more realistic. Abdulatif et al. \cite{Abdulatif2024_Arxiv} extended CMGAN with ablation studies on model inputs and architectural design choices, demonstrating its ability to generalize to unseen noise types and distortions. Dai et al. used the CMGAN in speech enhancement with phone-fortified perceptual loss \cite{Dai2023}. Additionally, recent work by Ting Zhang explored the use of CMGAN for speech enhancement \cite{Zhang2024}. Belz et al. introduced a U-Net model that integrates multiple feature extraction stages, using convolutional components to capture different aspects of the audio signal while preserving important details \cite{Belz2024}. A recent development is the integration of gated convolutional mechanisms into U-Net, as proposed by Baloch et al. [13]. In their work, they developed a fully convolutional U-Net combined with a Gated Convolutional Neural Network (GCNN) to enhance speech quality by suppressing background noise and improving intelligibility \cite{Baloch2023}. This modification has proven effective in improving speech enhancement for both English and Urdu speech, emphasizing the adaptability of U-Net models for multilingual applications. A variational U-Net architecture was also introduced by Nustede et al. \cite{Nustede2021}. This model incorporates a probabilistic bottleneck into the U-Net structure, which captures intricate spectro-temporal relationships in the audio signal, thereby enhancing the model's generalizability across different noise conditions. This approach demonstrates how probabilistic modifications to U-Net can address some limitations of traditional deterministic models. Another development was done by Hossain et al. \cite{UNET2024} as a supervised single-channel speech enhancement by U-Net.

While these previous works primarily evaluated models using datasets similar or identical to their training datasets, our research differs significantly in its methodology and evaluation strategy. As summarized in Table I, previous studies primarily evaluate speech enhancement models using the same datasets for training and evaluation, often under controlled conditions. In contrast, our study systematically compares three state-of-the-art models—Wave-U-Net, CMGAN, and U-Net—by retraining them using their recommended configurations but incorporating a common dataset (DEMAND) to introduce natural noise conditions consistently across all training scenarios. Moreover, we evaluate the trained models on completely independent, real-world datasets that differ significantly from those used during training, specifically the Clarkson dataset (child speakers aged 4–18), VPQAD dataset (adult speakers aged 18–40), and SpEAR dataset (artificially generated noise). Additionally, we uniquely integrate biometric speaker recognition evaluation (VeriSpeak) alongside traditional metrics (SNR and PESQ) to investigate the practical impact of enhancement techniques on voice biometrics applications, thus distinguishing our work from prior studies.

\section{Dataset Description}

\subsection{Training Datasets}
The datasets used for training the denoising model are as follows:

\subsubsection{DEMAND Dataset}

The DEMAND (Diverse Environments Multichannel Acoustic Noise Database) dataset is a comprehensive collection of real-world noise recordings captured in various environments, including domestic, office, and outdoor settings. Each recording uses 16 microphones to capture the spatial characteristics, which refer to the properties of sound or noise that describe how it behaves in a physical space, providing a more realistic representation of acoustic conditions \cite{DEMAND2013}.

\subsubsection{MUSDB18 HQ Dataset}

The MUSDB18 HQ dataset is an extended version of the original MUSDB18 dataset, providing high-quality multi-track audio recordings for music source separation tasks. Unlike the standard MUSDB18, the HQ version offers higher-fidelity audio with detailed temporal and spectral information \cite{MUSDB2019}. The dataset consists of separate stems for instruments such as vocals, bass, drums, and other accompaniment. 

In our study, we utilized the MUSDB18 HQ dataset. We introduced different types of environmental noise from the DEMAND dataset into the other stem to simulate challenging real-world conditions for training the Wave-U-Net model.

\subsubsection{VCTK Corpus Dataset}

The VCTK Corpus is a multi-speaker English speech dataset produced by the Centre for Speech Technology Research (CSTR) at the University of Edinburgh \cite{CSTR2019}. It consists of clean speech recordings from 110 speakers with diverse accents. In this study, we used the VCTK dataset as a source of clean speech, which was subsequently mixed with noise from the DEMAND dataset to create challenging noisy conditions for training the CMGAN model.

\subsubsection{Librispeech Corpus Dataset}

The LibriSpeech Corpus is a large-scale corpus dataset of read English speech designed for automatic speech recognition (ASR) tasks. It was introduced by Panayotov et al. \cite{Panayotov2015} and is derived from public-domain audiobooks. The dataset contains approximately 1,000 hours of high-quality speech sampled at 16 kHz. In this study, the LibriSpeech dataset was used as a source of clean speech data, providing a foundation for training the U-Net model.

\subsubsection{ESC-50 Dataset}
The ESC-50 dataset is an environmental sound classification dataset introduced by Piczak \cite{Piczak2015}. It has 2,000 audio recordings across 50 diverse classes, including alarms, natural sounds, human voices, and mechanical noises. Each class contains 40 recordings, and the dataset is recorded at a sampling rate of 44.1 kHz, providing high-quality audio suitable for evaluating noise classification and speech enhancement models. In this study, the ESC-50 dataset was used to create challenging noisy conditions by mixing it with the DEMAND dataset, contributing to the training of the U-Net variant.
\\
Table 2 shows the characteristics of the training datasets used in this study, including the number of files, audio type, sampling rate, and audio format.

\begin{table}[htbp]
\centering
\small
\caption{Summary of Training Datasets Used for Model Development}
\label{tab:training_datasets}
\renewcommand{\arraystretch}{1.5} 
\setlength{\tabcolsep}{1.5pt} 
\begin{tabular}{>{\centering\arraybackslash}p{2cm}|
                >{\centering\arraybackslash}p{1.8cm}|
                >{\centering\arraybackslash}p{2cm}|
                >{\centering\arraybackslash}p{1.3cm}|
                >{\centering\arraybackslash}p{1cm}}
\hline
\textbf{Dataset}           & \textbf{Number of Files/Length} & \textbf{Audio Type}        & \textbf{Sampling Rate} & \textbf{Format} \\ \hline
DEMAND Dataset             & 288; 24 hours                              & Real-world noise           & 48 kHz                 & .wav                  \\ \hline
MUSDB18 HQ Dataset           & 150; 9+ hours                             & Multi-track (vocals, bass, drums, etc.) & 44.1 kHz              & .wav                  \\ \hline
VCTK Corpus Dataset        & 44,000+; 300+ hours                         & Clean speech               & 48 kHz                 & .flac                 \\ \hline
Librispeech Corpus Dataset & 1,000 hours                     & Read speech                & 16 kHz                 & .flac                  \\ \hline
ESC 50 Dataset             & 2,000; 2.78 hours                           & Environmental sounds       & 44.1 kHz               & .wav                  \\ \hline
\end{tabular}
\end{table}

\subsection{Evaluation  Datasets}

To assess the model performance under diverse and challenging real-world conditions, the following evaluation datasets were in this study:

\subsubsection{SpEAR Dataset}

The SpEAR dataset (Speech Enhancement Evaluation with Added Random Noise) is designed to evaluate the robustness of speech enhancement models under varying noise conditions. It contains speech recordings to which Gaussian noise has been manually added at different levels, creating a wide spectrum of noisy conditions for rigorous testing. This dataset consists of 70 files, each with a unique noise level, and it is generally considered text-independent and recorded in WAV format to ensure high-quality audio suitable for both subjective and objective evaluations \cite{SpEAR2006}. This study utilized the SpEAR dataset to assess the effectiveness of the Wave-U-Net, CMGAN, and U-Net models in enhancing speech quality in challenging noise scenarios.

\subsubsection{VPQAD Dataset}

The VPQAD (Voice Pre-Processing and Quality Assessment Dataset) is a comprehensive dataset designed to evaluate speech quality in noisy, real-world environments such as cafeterias and laboratory environments. It contains voice recordings from 50 participants aged 18 to 40 in text-dependent and independent format \cite{VPQAD2024}. This study used VPQAD as an evaluation dataset to test how well the models generalize to unpredictable real-world conditions.

\subsubsection{Clarkson Dataset}

The Clarkson Dataset is a unique voice dataset collected from children aged 4 to 18 over nine years. The latest update includes 1,656 recordings, collected approximately every six months starting from 2016. Each recording lasted around 90 seconds and was gathered from local schools in a controlled but non-soundproof environment, resulting in natural background noises such as doors opening and people talking. The dataset captures text-dependent tasks and is not publicly available due to privacy restrictions. Fig. 1 illustrates the data collection setup, showing a participant during recording, with two different types of microphones indicated.

\begin{figure}[htbp] 
    \centering
    \includegraphics[width=0.9\linewidth]{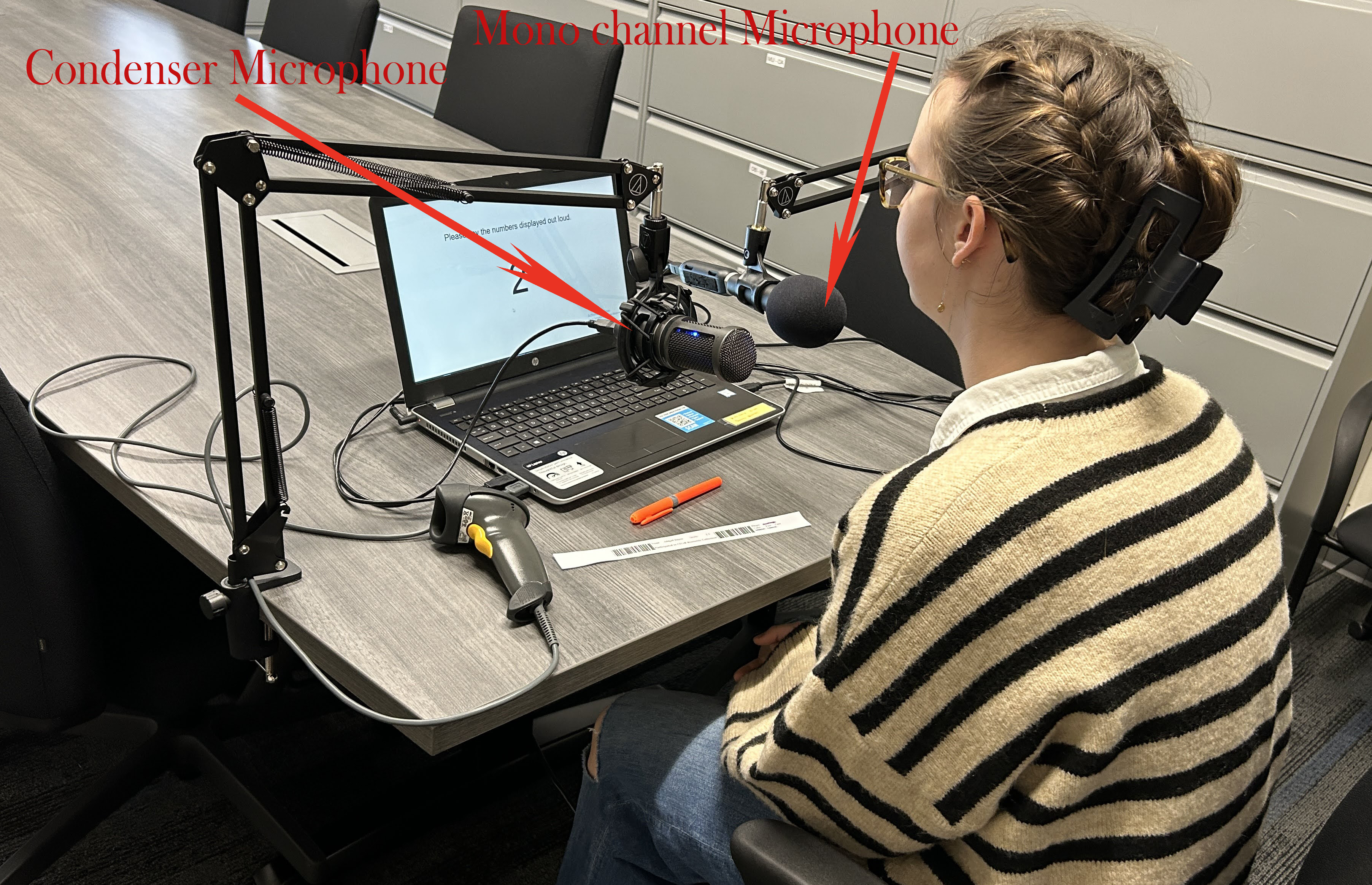}  
    \caption{Data Collection Setup for the Clarkson Dataset, Showing Participant with Dual Microphone Setup}
    \label{fig:clarkson_dataset_setup} 
\end{figure}

The Clarkson Dataset, with its focus on age-related changes in children's voices, offers valuable insights for vocal development and robust speech recognition \cite{Purnapatra2020} \cite{Alim2024}.

\begin{table}[htbp]
\centering
\small
\caption{Summary of Evaluation Datasets for Model Performance Evaluation}
\label{tab:evaluation_datasets}

{\fontsize{8}{9}\selectfont
\label{tab:sub_pairs_comparison3}

\renewcommand{\arraystretch}{1.5} 
\setlength{\tabcolsep}{1.5pt} 
\begin{tabular}{>{\centering\arraybackslash}p{1cm}|
                >{\centering\arraybackslash}p{1.6cm}|
                >{\centering\arraybackslash}p{1.4cm}|
                >{\centering\arraybackslash}p{1cm}|
                >{\centering\arraybackslash}p{1.8cm}|
                >{\centering\arraybackslash}p{1.3cm}}
\hline
\textbf{Dataset}       & \textbf{Number of Files/Subjects} & \textbf{Noise Type}       & \textbf{Sampling Rate} & \textbf{Recording Environment}           & \textbf{Availability}     \\ \hline
SpEAR Dataset          & 70 files, 7 subjects, 2 hours (Approx.)             & Gaussian noise (various levels) & 16 kHz                & Controlled, clean speech mixed with noise & Publicly Available        \\ \hline
VPQAD Dataset          & 50 participants, 2+ hours                  & Real-world noise (various environments) & 41 kHz         & Controlled real-life environments         & Publicly Available        \\ \hline
Clarkson Dataset       & 1,656 files, 184 (avg.) participants, 32+ hours  & Real-world ambient noises & 41 kHz                & Non-soundproof, local schools             & Not Publicly Available    \\ \hline
\end{tabular}
}
\end{table}

Table 3 provides an overview of the evaluation datasets used in this study, including key attributes such as the number of files, noise type, sampling rate, recording environment, and availability.

\section{Methodology}

This study begins with dataset preparation, followed by model training details. The trained models are then assessed using objective metrics to determine their effectiveness in enhancing speech quality and robustness in diverse conditions. All codes pertaining to model architectures, data pre-processing, testing, training and calculation of evaluation metrics are available at the following GitHub repository: \href{https://github.com/jahangirkhondkar/DL_SpeechEnhancementToolkit}{DL Speech Enhancement Toolkit}

\subsection{Model Architectures}

\subsubsection{Wave-U-Net Architecture}
Wave-U-Net is a fully convolutional neural network designed to process raw audio signals directly in the time domain. It uses a U-Net architecture to perform end-to-end speech enhancement and source separation. The model consists of an encoder-decoder structure with skip connections to preserve fine-grained information throughout the network, making it highly effective for speech enhancement tasks. Figure 2 provides a detailed schematic of the Wave-U-Net model used in this study.

\begin{figure}[htbp] 
    \centering
    \includegraphics[width=1\linewidth]{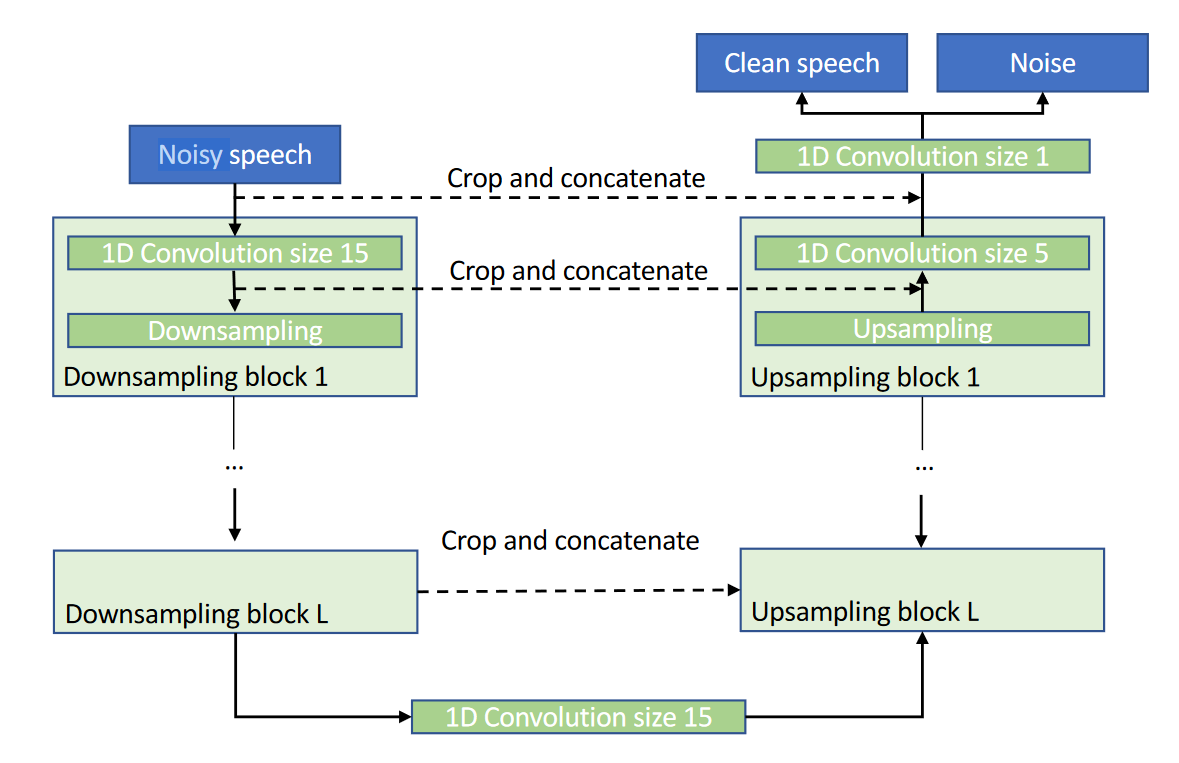}  
    \caption{The Wave-U-Net Architecture \cite{Stoller2018} \cite{Macartney2018}}
    \label{fig: WaveuNet_architecture} 
\end{figure}

\textbf{Encoder (Down-Sampling Path):} The encoder consists of 1D convolutional layers that progressively reduce the temporal resolution while increasing the number of feature maps, thereby capturing high-level representations of the input audio signal.

\textbf{Skip Connections:} Skip connections link each encoder layer to the decoder's corresponding layer, allowing detailed features from the input to be preserved and reused during reconstruction.

\textbf{Decoder (Up-Sampling Path):} The decoder uses transposed convolutional layers to restore the signal's temporal resolution. It combines information from skip connections to reconstruct a denoised output, effectively enhancing clean speech while reducing noise.

\subsubsection{CMGAN Architecture}

The Conformer-Based Metric-GAN (CMGAN) architecture is designed for monaural speech enhancement. It employs a combination of generative adversarial networks (GAN) and conformers to improve speech quality. CMGAN consists of a generator and a metric discriminator, each addressing different aspects of the enhancement process. Figure 3 presents the schematic of the CMGAN model.

\begin{figure}[htbp] 
    \centering
    \includegraphics[width=1\linewidth]{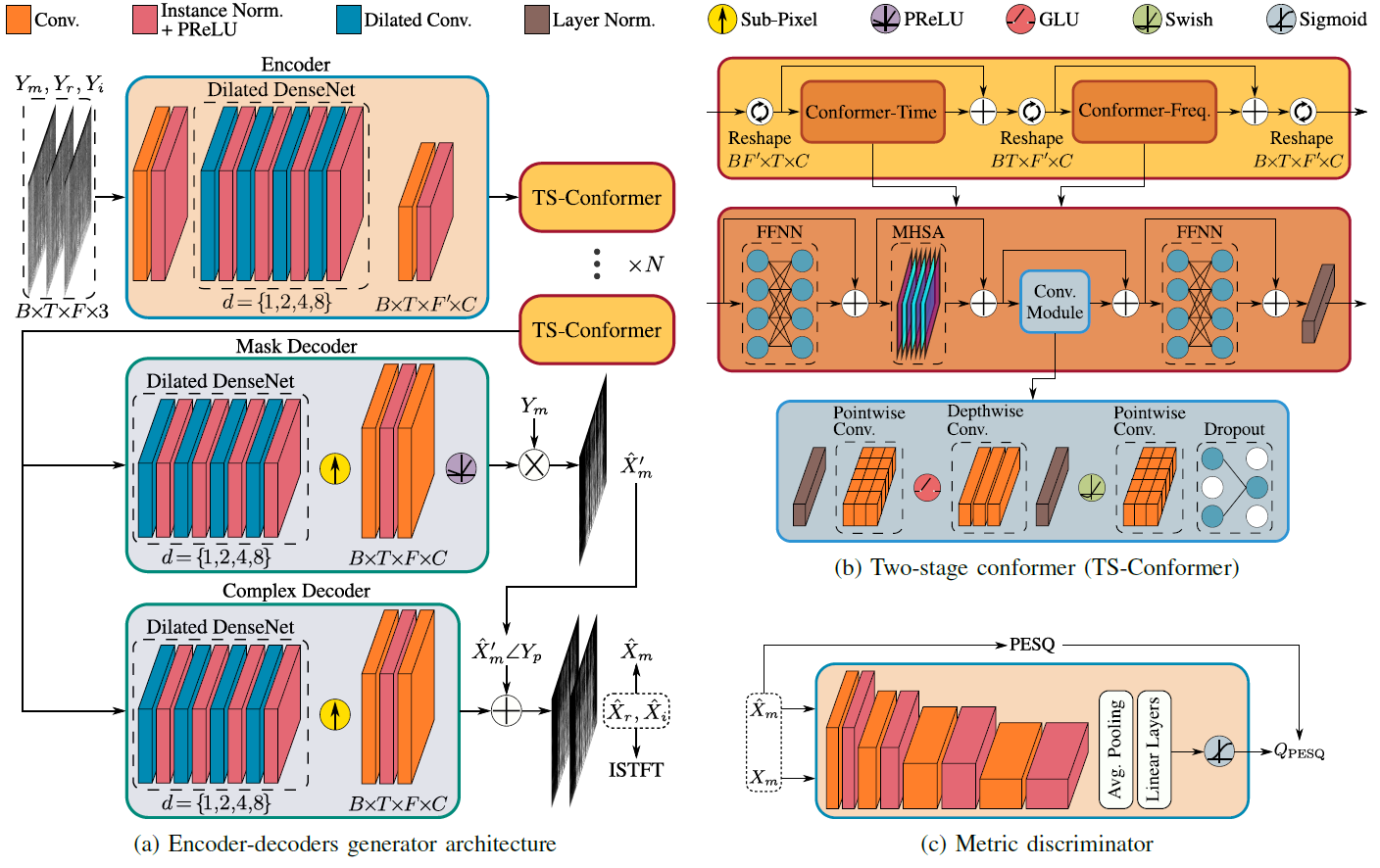}  
    \caption{Overview of the CMGAN Architecture \cite{Abdulatif2024_Arxiv}}
    \label{fig: CMGAN_architecture} 
\end{figure}

\textbf{Generator:} The generator employs a shared encoder that combines the speech signal's magnitude and complex components (real and imaginary parts). It integrates two-stage conformer blocks to capture both temporal and frequency dependencies effectively. The generator then decouples into two pathways: one for magnitude masking and the other for refining the real and imaginary parts.

\textbf{Metric Discriminator:} The discriminator estimates the PESQ score by learning from the clean and enhanced spectrum. This guides the generator in producing higher-quality outputs, helping optimize non-differentiable quality metrics directly and making the enhanced speech sound more natural.

\subsubsection{U-Net Architecture}
U-Net is a fully convolutional neural network initially designed for biomedical image segmentation. It features a symmetric encoder-decoder structure, with a contracting path (encoder) to capture context and a symmetric expanding path (decoder) to localize, as shown in Figure 4.

\begin{figure}[htbp] 
    \centering
    \includegraphics[width=1\linewidth]{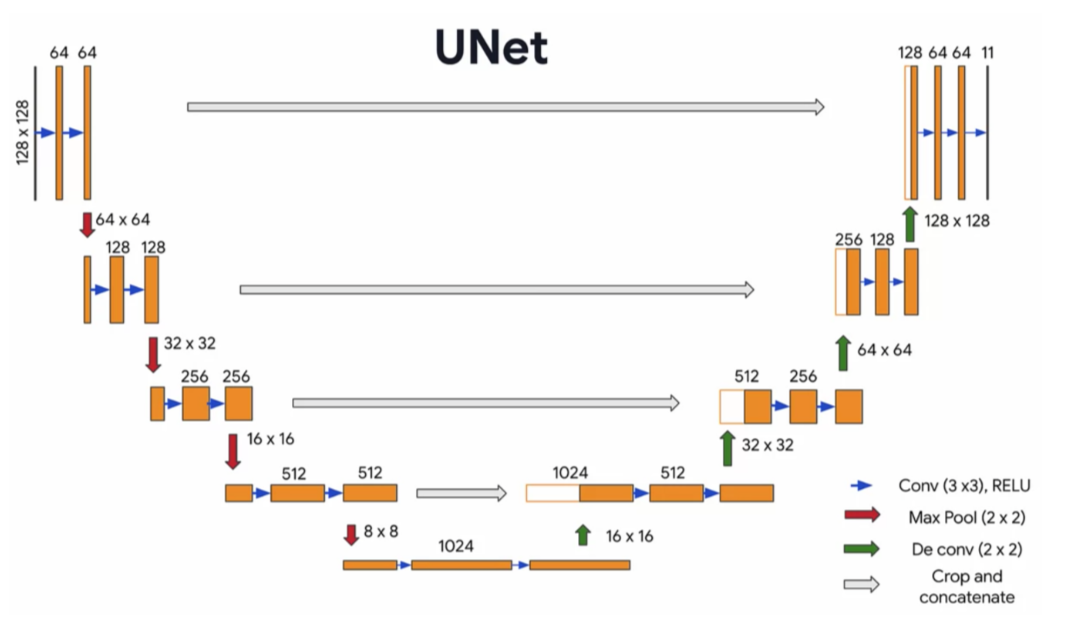}  
    \caption{Overview of the U-Net Architecture \cite{UNet2015}}
    \label{fig: UNet_architecture} 
\end{figure}

The encoder consists of repeated 3x3 convolutions followed by ReLU activation and 2x2 max pooling for down-sampling, doubling the number of feature channels at each step. The decoder up-samples the feature maps using transposed convolutions, combining them with high-resolution features from the encoder through skip connections. This combination helps the network preserve fine details while learning contextual features, which makes it particularly effective for speech enhancement and similar tasks.

\subsection{Data Preprocessing}

Preprocessing steps were taken for each of the three models—Wave-U-Net, CMGAN, and U-Net—before training and evaluation. Different preprocessing approaches were required due to variations in dataset formats and model requirements.

\subsubsection{Wave-U-Net:}

\textbf{Training Preprocessing:} The MUSDB18 HQ dataset used for training had a sampling rate of 44.1 kHz, which matches the input requirement of Wave-U-Net. However, the noise signals added from the DEMAND dataset were originally recorded at 48 kHz. Therefore, all tracks from the DEMAND dataset were downsampled to 44.1 kHz.

\textbf{Enhancement Preprocessing:} The Clarkson and VPQAD datasets were both at 44.1 kHz, and the SpEAR dataset had a sampling rate of 16 kHz. Therefore, upsampling to 44.1 kHz was performed to make it compatible with Wave-U-Net's trained model during evaluation.

\subsubsection{CMGAN}

\textbf{Training Preprocessing:} The VCTK Corpus dataset, initially in .flac format and sampled at 48 kHz, was converted to .wav format. To match the target sampling rate of the CMGAN model, the clean speech files were downsampled to 16 kHz. The DEMAND dataset was similarly downsampled to 16 kHz and mixed with the VCTK dataset to generate noisy speech samples for training.

\textbf{Enhancement Preprocessing:} CMGAN checkpoints were configured for 16 kHz audio, necessitating downsampling of Clarkson and VPQAD datasets to 16 kHz before evaluation. The SpEAR dataset was already at 16 kHz, so it was directly compatible with CMGAN for enhancement.

\subsubsection{U-Net}
\textbf{Training Preprocessing:} The LibriSpeech Corpus dataset, sampled at 16 kHz, was originally in .flac format and converted to .wav format to be compatible with the U-Net model. The ESC-50 dataset, recorded at 44.1 kHz, was downsampled to 16 kHz and mixed with the LibriSpeech dataset.

\textbf{Enhancement Preprocessing:} Since U-Net checkpoints were compatible only with 16 kHz audio, Clarkson and VPQAD datasets were downsampled to 16 kHz for evaluation. The SpEAR dataset, at 16 kHz, was compatible without additional preprocessing.
\\

Wave-U-Net follows Stoller et al. \cite{Stoller2018}, whose original MUSDB18-HQ training data are 44.1 kHz; keeping this rate preserves the network’s time-domain receptive field and avoids aliasing from down-sampling. CMGAN and U-Net, which employ STFT front-ends, are trained at the standard 16 kHz speech rate. After enhancement, all outputs are resampled to 16 kHz so that PESQ, SNR, and VeriSpeak remain directly comparable across models.

\subsection{Training Setup}

\subsubsection{Training Environment}
All model training was conducted on a local system with an NVIDIA GeForce RTX 4080 GPU, featuring a total memory of 16,376 MiB, with a Compute Capability 8.9. The system ran on Ubuntu 22.04, utilizing 64 GB of RAM and an AMD Processor. Python virtual environments were created for each model with all required libraries were installed. MATLAB 2022b and VSCode were also used for parts of the evaluation and data processing tasks.

\subsubsection{Training parameters}

The training parameters varied across the three models—Wave-U-Net, CMGAN, and U-Net—based on their specific architecture requirements and dataset properties. While the initial training parameters were informed by the configurations provided by the original authors in their GitHub repositories, we primarily retained these default parameters as our study used the same datasets on which their models were trained. We added common noisy files from the CMGAN dataset to introduce slight diversity while maintaining the original format and structure. This approach ensured consistency with the original setup, allowing fair comparisons, while small adjustments, such as checkpointing, optimized training for our modified datasets. The details of each model's training configuration are provided in the following subsections.
\\
\\
\textbf{Wave-U-Net:}

The Wave-U-Net model was trained using the Adam optimizer, with an initial learning rate of 0.0001. A cyclic learning rate schedule was employed, oscillating between 1e-3 and 5e-5 to help the model avoid local minima. The loss functions used were Mean Squared Error (MSE) and L1 loss, which aided in preserving sharp transients in the audio signal. The training was conducted with a batch size of 16. Additionally, data augmentation was applied, involving random amplification between 0.7 and 1.0 to enhance the model's generalization capabilities.
\\
\\
\textbf{CMGAN:}

The CMGAN model was trained using the AdamW optimizer for both the generator and discriminator. The initial learning rates were set as follows:
\[
\text{The learning rate for the generator is } 5 \times 10^{-4}.
\]
\[
\text{The learning rate for the discriminator is } 1 \times 10^{-3}.
\]

A cyclic learning rate scheduler was employed, with decay by a factor of 0.5 every 12 epochs. The batch size was set to 4, ensuring efficient training while maintaining computational feasibility.

The training loss consisted of a combination of time-frequency loss and adversarial loss, balanced using the following weight factors:

\[
\gamma_1 = 1, \quad \gamma_2 = 0.01, \quad \gamma_3 = 1,
\]

where:
\begin{itemize}
    \item \(\gamma_1\) represents the weight of the time-domain loss,
    \item \(\gamma_2\) represents the weight of the frequency-domain loss,
    \item \(\gamma_3\) represents the weight of the adversarial loss.
\end{itemize}

\textbf{U-Net:}

The U-Net model was trained using the Adam optimizer, with an initial learning rate 0.0002. A learning rate scheduler was applied to reduce the learning rate by half after every 20 epochs to stabilize training and improve convergence. The model used a batch size of 16 for efficient training, and the L1 loss function was used to minimize the difference between the predicted and clean speech signals. Data augmentation was performed through random noise addition and pitch shifts to improve the model's robustness to different audio conditions \cite{Belz2024}.

\subsubsection{Experimental Setup}

\textbf{Wave-U-Net:}

The MUSDB18-HQ dataset consists of two folders: one containing a training set ``train" of 100 songs and another with a test set ``test" of 50 songs. We used 120 folders for training and 30 for validation, ensuring an 80-20 split for model evaluation. In the ``others" stem of the training folders, 100 folders were randomly replaced with noise from the DEMAND dataset to simulate more challenging training conditions. The training process ran for 68 epochs, and the model checkpoint with the lowest validation loss (0.0248) was selected as the final model. Early stopping was employed, halting training if there was no improvement in validation loss for 20 consecutive epochs. Validation loss was monitored throughout training using TensorBoard, and a graph showing validation loss versus epoch is presented in the results section to illustrate model convergence.

\textbf{CMGAN:}

The experiment involved mixing the Voice Bank Corpus dataset with noise from the DEMAND dataset, creating a dataset with 10 to 15 dB SNR variation. A total of 2000 audio tracks were generated by mixing clean audio with random noises from the DEMAND dataset, which were then split into 1600 for training and 400 for validation to maintain an 80-20 split. The training process was conducted over 120 epochs, and the best checkpoint, based on validation losses, was found at the 34th epoch, which was used for audio enhancement.

\textbf{U-Net:}

The experiment used 2000 files from the LibriSpeech Corpus as clean speech and a combination of 288 files from the DEMAND dataset and 1712 files from the ESC-50 dataset as noise sources. The files were divided into 1,600 for training and 400 for validation for each dataset, maintaining an 80-20 split for training and evaluation. The training ran for 100 epochs, with the model checkpoint saved based on the lowest validation loss. After careful validation, the best model was selected, providing optimal performance in enhancing noisy speech. The evaluation included standard objective metrics like PESQ and SNR to assess the effectiveness of the enhanced audio.

\subsection{Evaluation Metrics}

\subsubsection{ Signal-to-Noise Ratio (SNR)}

The signal-to-noise ratio (SNR) quantifies the clarity of a speech signal by comparing the level of the desired signal to the level of background noise. In speech enhancement, a higher SNR value indicates that the enhancement process has effectively reduced noise, resulting in a clearer signal \cite{Proakis2001}. SNR is typically measured in decibels (dB), calculated as:
\[
\text{SNR} = 10 \log_{10} \left( \frac{P_{\text{signal}}}{P_{\text{noise}}} \right)
\]

Where:
\begin{itemize}
    \item \(P_{\text{signal}}\) is the power of the signal.
    \item \(P_{\text{noise}}\) is the power of the noise.
\end{itemize}

A positive SNR indicates that the signal power is greater than the noise power, while a negative SNR implies the noise dominates the signal.

The SNR script uses STFT to compute signal and noise power for SNR calculation, with a 32 ms frame length (512 samples at 16 kHz) and 50\% overlap. Noise is estimated from the first 10\% of the unfiltered audio signal, under the assumption that this portion contains background noise and minimal speech content. This segment is used as a reference for noise power calculation in the SNR computation process \cite{Zhao2016} \cite{SNR2024}.

\subsubsection{Perceptual Evaluation of Speech Quality (PESQ)}

PESQ is an objective method standardized by the International Telecommunication Union (ITU-T Recommendation P.862) \cite{PESQ2024} for assessing the quality of speech signals in telecommunication systems. In this study, we used the PESQ library \cite{PESQPython} available in the Python environment. PESQ evaluates the perceived quality of a speech signal by comparing it with a reference. We supplied noisy audio as the degraded version and filtered audio as the reference. PESQ measures perceptual differences (e.g., noise reduction, distortion) and outputs a score, typically ranging from -0.5 to 4.5, with higher scores indicating better perceived quality \cite{PESQ2001}.

\subsubsection{VeriSpeak: Speaker Recognition Performance Evaluation}

 VeriSpeak, developed by Neurotechnology \cite{VeriSpeak2024}, is a commercial tool widely used for biometric speaker verification. It can match voice samples to enrolled voice templates by calculating similarity scores. This software was chosen to assess how well the enhancement process retained speaker-specific features. High scores indicate effective retention of the speaker's identity, while low scores suggest possible distortion during enhancement. A MATLAB script was written to automate the enrollment and verification of voice samples, with a False Acceptance Rate (FAR) of 0.01\%, and the results were logged in Excel for further analysis.

\section{ Results}

We analyzed three datasets using three evaluation metrics across three models. Due to a large number of files, we have presented only a subset of results evaluated. 

\subsection{SpEAR Dataset Result:}

The results of SNR, PESQ, and VeriSpeak for the SpEAR dataset for the Wave-U-Net, CMGAN, and U-Net models are as follows-
\\

\begin{table}[htbp]
\centering
\small
\caption{Combined SNR Changes for the SpEAR Dataset across All Models}
\label{tab:performance_comparison1}
\renewcommand{\arraystretch}{1} 
\setlength{\tabcolsep}{1.8pt}   
{\fontsize{7.5}{8.5}\selectfont
\begin{tabular}{p{1cm}|c|c|c|c|c|c|c|c|c}
\hline
\textbf{Subjects} & \multicolumn{3}{c|}{\textbf{Wave-U-Net}} & \multicolumn{3}{c|}{\textbf{CMGAN}} & \multicolumn{3}{c}{\textbf{U-Net}} \\ \hline
                  & \textbf{Unf} & \textbf{Fil} & \textbf{Chng\%} & \textbf{Unf} & \textbf{Fil} & \textbf{Chng\%} & \textbf{Unf} & \textbf{Fil} & \textbf{Chng\%} \\ \hline
bigtips..    & 19.72      & 18.56       & -5.87          & 19.72      & 18.29      & -7.27          & 19.72      & 50.95      & 158.32         \\ \hline
b.burst.. & 34.15   & 16.68       & -51.15         & 34.15      & 28.38      & -16.90         & 34.15      & 50.99      & 49.30          \\ \hline
b.f16r..   & 31.87   & 22.53       & -29.32         & 31.87      & 26.44      & -17.04         & 31.87      & 48.89      & 53.38          \\ \hline
b.facto..       & 26.68   & 21.23       & -20.43         & 26.68      & 23.29      & -12.73         & 26.68      & 49.74      & 86.40          \\ \hline
b.pinkr..  & 35.45   & 25.76       & -27.32         & 35.45      & 28.52      & -19.54         & 35.45      & 48.38      & 36.48          \\ \hline
b.volvor.. & 21.59   & 18.77       & -13.06         & 21.59      & 19.78      & -8.35          & 21.59      & 50.41      & 133.55         \\ \hline
b.whiter.. & 37.63   & 24.64       & -34.51         & 37.63      & 29.90      & -20.53         & 37.63      & 48.21      & 28.11          \\ \hline
butter..         & 21.32   & 19.87       & -6.77          & 21.32      & 19.50      & -8.51          & 36.03      & 49.72      & 38.02          \\ \hline
bu.burstr..  & 36.03   & 13.52       & -62.48         & 36.03      & 30.56      & -15.16         & 31.52      & 48.45      & 53.70          \\ \hline
bu.f16r..    & 31.52   & 21.25       & -32.59         & 31.52      & 26.49      & -15.95         & 26.93      & 49.12      & 82.37          \\ \hline
Avg. for the above files & 29.59 & 20.28  & \textbf{-28.35\%} & 29.38   & 24.96      & \textbf{-14.20\%} & 30.16  & 49.48      & \textbf{71.96\%} \\ \hline
Avg. for all files in dataset & 32.79 & 19.82 & \textbf{-33.48\%} & 30.45 & 22.98 & \textbf{-23.25\%} & 33.57 & 49.35 & \textbf{61.44\%} \\ \hline
\end{tabular}
}
\end{table}

Table IV presents the percentage increases or decreases in the signal-to-noise ratio (SNR) for the SpEAR dataset across all three models: Wave-U-Net, CMGAN, and U-Net.

\begin{table}[htbp]
\centering
\small
\caption{Combined PESQ Scores for SpEAR Dataset across All Models}
\label{tab:pesq_scores_comparison}
\renewcommand{\arraystretch}{1} 
\setlength{\tabcolsep}{4.5pt}     
\begin{tabular}{c|c|c|c}
\hline
\textbf{Subjects} & \textbf{Wave-U-Net} & \textbf{CMGAN} & \textbf{U-Net} \\ \hline
bigtipsr1\_16       & 2.18               & 3.76           & 1.08           \\ \hline
bigtips\_burstr1\_16 & 2.27               & 4.14           & 1.29           \\ \hline
bigtips\_f16r1\_16   & 2.28               & 4.07           & 1.28           \\ \hline
bigtips\_factoryr1\_16 & 2.38             & 3.99           & 1.10           \\ \hline
bigtips\_pinkr1\_16  & 2.58               & 4.05           & 1.32           \\ \hline
bigtips\_volvor1\_16 & 2.64               & 3.83           & 1.10           \\ \hline
bigtips\_whiter1\_16 & 2.67               & 4.08           & 1.12           \\ \hline
butterr1\_16         & 2.82               & 4.05           & 1.06           \\ \hline
butter\_burstr1\_16  & 3.28               & 4.33           & 1.07           \\ \hline
butter\_f16r1\_16    & 3.76               & 4.12           & 1.03           \\ \hline
Avg. for above files         & \textbf{2.68}  & \textbf{4.04}  & \textbf{1.14}  \\ \hline
Avg. for all files in dataset         & \textbf{2.91}  & \textbf{3.33}  & \textbf{1.15}  \\ \hline
\end{tabular}
\end{table}

Table V summarizes the PESQ (Perceptual Evaluation of Speech Quality) scores, and Table VI provides the VeriSpeak scores for the SpEAR dataset for all three models.

\begin{table}[htbp]
\centering
\small
\caption{Combined VeriSpeak Scores for SpEAR Dataset across All Models}
\label{tab:comparison_unf_fil_20}
\renewcommand{\arraystretch}{1} 
\setlength{\tabcolsep}{2.5pt}     
{\fontsize{7.5}{8.5}\selectfont
\begin{tabular}{p{1cm}|p{0.7cm}|p{0.6cm}|c|c|c|p{0.7cm}|p{0.7cm}|c|p{0.7cm}}
\hline
\textbf{Subjects}         & \multicolumn{3}{c|}{\textbf{Wave-U-Net}} & \multicolumn{3}{c|}{\textbf{CMGAN}} & \multicolumn{3}{c}{\textbf{U-Net}} \\ \hline
                          & \textbf{Unf} & \textbf{Fil} & \textbf{Chng\%} & \textbf{Unf} & \textbf{Fil} & \textbf{Chng\%} & \textbf{Unf} & \textbf{Fil} & \textbf{Chng\%} \\ \hline
d.pinkr          & 108.5        & 115          & 5.99            & 108.5        & 106.5        & -1.84           & 108.5        & 0            &                 \\ \hline
d.whiter         & 89.5         & 102          & 13.97           & 89.5         & 91           & 1.68            & 89.5         & 0            &                 \\ \hline
s.burstr1     & 101          & 118          & 16.83           & 101          & 110.5        & 9.41            & 101          & 0            &                 \\ \hline
s.fac         & 83.5         & 78           & -6.59           & 83.5         & 91.5         & 0               & 83.5         & 0            &                 \\ \hline
s.pinkr          & 100.5        & 111.5        & 10.95           & 100.5        & 117.5        & 16.92           & 100.5        & 0            &                 \\ \hline
s.vol             & 91           & 104          & 14.29           & 91           & 88           & -3.30           & 91           & 0            &                 \\ \hline
s.white       & 77           & 89.5         & 16.23           & 77           & 99           & 28.57           & 77           & 0            &                 \\ \hline
scholarsr1           & 79.5         & 89.5         & 12.58           & 79.5         & 84           & 5.66            & 79.5         & 0            &                 \\ \hline
s.f16r1       & 66           & 68.5         & 3.79            & 66           & 0            & 0               & 66           & 0            &                 \\ \hline
drawr1                & 76           & 91.5         & 20.39           & 76           & 76.5         & 0.66            & 76           & 0            &                 \\ \hline
Avg. for above files      & 87.25        & 96.75        & \textbf{10.84\%} & 87.25        & 96.06        & \textbf{5.77\%}  &              &              &                 \\ \hline
Avg. for all files in dataset & 88.61    & 97.97        & \textbf{11.63\%} & 88.61        & 95.32        & \textbf{8.9\%}   &              &              &                 \\ \hline
\end{tabular}
}
\end{table}

\subsection{VPQAD Dataset Result:}

The results of SNR, PESQ, and VeriSpeak for the VPQAD dataset for the Wave-U-Net, CMGAN, and U-Net models are as follows-

Table VII presents the percentage change in SNR for the VPQAD dataset across the three models.

\begin{table}[htbp]
\centering
\small
\caption{Combined SNR Change for VPQAD Dataset across All Models}
\label{tab:performance_comparison_21}
\renewcommand{\arraystretch}{1} 
\setlength{\tabcolsep}{1.5pt}     
{\fontsize{7.5}{8.5}\selectfont
\begin{tabular}{p{1.3cm}|c|c|c|c|c|c|c|c|c}
\hline
\textbf{Subjects} & \multicolumn{3}{c|}{\textbf{Wave-U-Net}} & \multicolumn{3}{c|}{\textbf{CMGAN}} & \multicolumn{3}{c}{\textbf{U-Net}} \\ \hline
                  & \textbf{Unf} & \textbf{Fil} & \textbf{Chng\%} & \textbf{Unf} & \textbf{Fil} & \textbf{Chng\%} & \textbf{Unf} & \textbf{Fil} & \textbf{Chng\%} \\ \hline
sub001     & 27.69        & 5.97         & -78.45          & 27.69        & 8.41         & -69.63          & 27.69        & 50.25        & 81.44           \\ \hline
sub002     & 27.57        & 4.00         & -85.48          & 27.57        & 10.45        & -62.09          & 27.57        & 50.48        & 83.05           \\ \hline
sub003     & 31.63        & 4.05         & -87.18          & 31.63        & 10.14        & -67.92          & 31.63        & 50.23        & 58.79           \\ \hline
sub004     & 31.43        & 5.24         & -83.33          & 31.43        & 11.88        & -62.18          & 31.43        & 49.82        & 58.52           \\ \hline
sub005     & 33.77        & 7.49         & -77.83          & 33.77        & 13.00        & -61.50          & 33.77        & 49.75        & 47.34           \\ \hline
sub006     & 32.15        & 2.78         & -91.35          & 32.15        & 12.86        & -59.98          & 32.15        & 49.70        & 54.61           \\ \hline
sub007     & 28.27        & 2.71         & -90.40          & 28.27        & 10.31        & -63.51          & 28.27        & 50.22        & 77.64           \\ \hline
sub008     & 28.84        & 4.47         & -84.50          & 28.84        & 11.29        & -60.84          & 28.84        & 50.44        & 74.89           \\ \hline
sub009     & 29.79        & 3.42         & -88.51          & 29.79        & 9.56         & -67.89          & 29.79        & 50.03        & 67.94           \\ \hline
sub010     & 34.44        & 4.47         & -87.02          & 34.44        & 13.88        & -59.69          & 34.44        & 49.60        & 44.04           \\ \hline
Avg. for above files  & 30.56        & 4.46         & \textbf{-85.41\%} & 30.56        & 11.18        & \textbf{-63.52\%} & 30.56        & 50.05        & \textbf{64.83\%}  \\ \hline
Avg. for all files in dataset  & 30.12        & 5.02         & \textbf{-83.33\%} & 30.12        & 10.40        & \textbf{-65.57\%} & 30.12        & 50.09        & \textbf{67.05\%}  \\ \hline
\end{tabular}
}
\end{table}

Table VIII summarizes the PESQ scores for the VPQAD dataset, providing a comparison of the three models' perceived speech quality improvements.

\begin{table}[htbp]
\centering
\small
\caption{Combined PESQ Scores for VPQAD Dataset across All Models}
\label{tab:model_scores_22}
\renewcommand{\arraystretch}{1} 
\setlength{\tabcolsep}{3.5pt}     
\begin{tabular}{c|c|c|c}
\hline
\textbf{Subjects} & \multicolumn{1}{c|}{\textbf{Wave-U-Net}} & \multicolumn{1}{c|}{\textbf{CMGAN}} & \multicolumn{1}{c}{\textbf{U-Net}} \\ \hline
sub001     & 1.23                                   & 1.38                                & 1.09                                \\ \hline
sub002     & 1.34                                   & 1.61                                & 1.28                                \\ \hline
sub003     & 1.19                                   & 1.40                                & 1.59                                \\ \hline
sub004     & 1.37                                   & 1.52                                & 1.48                                \\ \hline
sub005     & 1.24                                   & 1.32                                & 1.15                                \\ \hline
sub006     & 1.16                                   & 1.50                                & 1.43                                \\ \hline
sub007     & 1.39                                   & 1.61                                & 1.67                                \\ \hline
sub008     & 1.26                                   & 1.50                                & 1.36                                \\ \hline
sub009     & 1.19                                   & 1.37                                & 1.32                                \\ \hline
sub010     & 1.19                                   & 1.35                                & 1.18                                \\ \hline
Avg. for above files  & \textbf{1.26}                          & \textbf{1.46}                       & \textbf{1.36}                       \\ \hline
Avg. for all files in dataset  & \textbf{1.19}                          & \textbf{1.35}                       & \textbf{1.35}                       \\ \hline
\end{tabular}
\end{table}

Table IX shows the VeriSpeak match scores for the VPQAD dataset, comparing the three models' performance in preserving speaker identity after enhancement.

\begin{table}[htbp]
\centering
\small
\caption{Combined VeriSpeak Match Scores for VPQAD Dataset Across All Models}
\label{tab:performance_comparison_23}
\renewcommand{\arraystretch}{1} 
\setlength{\tabcolsep}{1.8pt}     
{\fontsize{7.5}{8.5}\selectfont
\begin{tabular}{p{1cm}|c|c|c|c|c|c|c|c|c}
\hline
\textbf{Subjects} & \multicolumn{3}{c|}{\textbf{Wave-U-Net}} & \multicolumn{3}{c|}{\textbf{CMGAN}} & \multicolumn{3}{c}{\textbf{U-Net}} \\ \hline
                  & \textbf{Unf} & \textbf{Fil} & \textbf{Chng\%} & \textbf{Unf} & \textbf{Fil} & \textbf{Chng\%} & \textbf{Unf} & \textbf{Fil} & \textbf{Chng\%} \\ \hline
sub001     & 81           & 95.5         & 17.90           & 81           & 114          & 40.74           & 81           & 0            &                 \\ \hline
sub002     & 65           & 83.5         & 28.46           & 65           & 101.5        & 56.15           & 65           & 165          & 153.85          \\ \hline
sub003     & 114.5        & 134.5        & 17.47           & 114.5        & 123.5        & 7.86            & 114.5        & 175.5        & 53.28           \\ \hline
sub004     & 120.5        & 133          & 10.37           & 120.5        & 0            & 0               & 120.5        & 171          & 41.91           \\ \hline
sub005     & 63.5         & 80.5         & 26.77           & 63.5         & 83.5         & 31.50           & 63.5         & 0            &                 \\ \hline
sub006     & 56           & 86.5         & 54.46           & 56           & 99.5         & 77.68           & 56           & 0            &                 \\ \hline
sub007     & 98.5         & 113.5        & 15.23           & 98.5         & 113          & 14.72           & 98.5         & 0            &                 \\ \hline
sub008     & 77           & 98.5         & 27.92           & 77           & 113          & 46.75           & 77           & 0            &                 \\ \hline
sub012     & 88.5         & 108.5        & 22.60           & 88.5         & 100.5        & 13.56           & 88.5         & 167          & 88.70           \\ \hline
sub013     & 77           & 117.5        & 52.60           & 77           & 120.5        & 56.49           & 77           & 0            &                 \\ \hline
Avg. for above files  & 84.15       & 105.15       & \textbf{27.38\%}  & 84.15       & 96.9       & \textbf{34.55\%}  &              &              &                 \\ \hline
Avg. for all files in dataset  & 91.57       & 117.52       & \textbf{30.22\%}  & 91.57       & 113.53       & \textbf{26.59\%}  &              &              &                 \\ \hline
\end{tabular}
}
\end{table}

\subsection{Clarkson Dataset Result:}

The Clarkson Dataset is more complex, with multiple collections, so presenting these results in separate tables makes sense. Each collection will have its own set of tables for SNR, PESQ, and VeriSpeak.\\

Tables X provide the SNR values for the Clarkson dataset, which includes five subjects, each from the five collections evaluated by the Wave-U-Net, CMGAN, and U-Net models and Table XI summarizes the PESQ (Perceptual Evaluation of Speech Quality) scores for the Clarkson dataset for the three models for five collections and ten subjects each.

\begin{table*}[htbp]
\centering
\small
\caption{SNR Results for Clarkson Dataset of 5 Collections for Wave-U-Net, CMGAN, and U-Net Models}
\label{tab:related_work_24}
\renewcommand{\arraystretch}{1.1} 
\setlength{\tabcolsep}{3.5pt} 
\begin{tabular}{c|c|c|c|c|c|c|c|c|c|c|c|c|c|c|c}
\hline
\textbf{Sub} & \multicolumn{3}{c|}{\textbf{20160104002}} & \multicolumn{3}{c|}{\textbf{20160104004}} & \multicolumn{3}{c|}{\textbf{20160104007}} & \multicolumn{3}{c|}{\textbf{20160104014}} & \multicolumn{3}{c}{\textbf{20160104015}} \\ \hline
             & \textbf{Unf} & \textbf{Fil} & \textbf{Chng\%} & \textbf{Unf} & \textbf{Fil} & \textbf{Chng\%} & \textbf{Unf} & \textbf{Fil} & \textbf{Chng\%} & \textbf{Unf} & \textbf{Fil} & \textbf{Chng\%} & \textbf{Unf} & \textbf{Fil} & \textbf{Chng\%} \\ \hline
\multicolumn{16}{|c|}{\textbf{Wave-U-Net}} \\ \hline
C6   & 9.20  & 2.57  & -72.05  & 19.78 & 8.67  & -56.16  & 16.03 & 10.16 & -36.61  & 13.62 & 4.84  & -64.46  & 16.47 & 7.51  & -54.41  \\ \hline
C7   & 23.78 & 14.85 & -37.55  & 15.78 & 3.60  & -77.19  & 29.13 & 17.25 & -40.78  & 17.65 & 6.94  & -60.68  & 24.63 & 13.82 & -43.89  \\ \hline
C8   & 21.21 & 11.96 & -43.61  & 15.99 & 5.00  & -68.73  & 14.35 & 5.39  & -62.44  & 21.14 & 13.11 & -37.98  & 15.92 & 5.68  & -64.32  \\ \hline
C13  & 11.23 & 3.27  & -70.88  & 12.53 & 4.93  & -60.65  & 14.39 & 5.34  & -62.89  & 18.17 & 5.05  & -72.21  & 14.12 & 6.30  & -55.38  \\ \hline
C14  & 13.57 & 3.62  & -73.32  & 12.01 & 3.33  & -72.27  & 13.15 & 3.70  & -71.86  & 11.84 & 2.06  & -82.60  & 12.95 & 3.30  & -74.52  \\ \hline
\multicolumn{16}{|c|}{\textbf{CMGAN}} \\ \hline
C6   & 9.20  & 4.63  & -49.65  & 19.78 & 16.1  & -18.58  & 16.03 & 11.96 & -25.38  & 13.62 & 9.05  & -33.55  & 16.47 & 12.1  & -26.73  \\ \hline
C7   & 23.78 & 18.30 & -23.04  & 15.78 & 10.92 & -30.80  & 29.13 & 22.77 & -21.83  & 17.65 & 11.46 & -35.07  & 24.63 & 20.22 & -17.90  \\ \hline
C8   & 21.21 & 10.20 & -51.91  & 15.99 & 7.19  & -55.03  & 14.35 & 6.35  & -55.75  & 21.14 & 14.19 & -32.88  & 15.92 & 7.26  & -54.40  \\ \hline
C13  & 11.23 & 5.83  & -48.09  & 12.53 & 7.94  & -36.63  & 14.39 & 7.81  & -45.73  & 18.17 & 11.09 & -38.97  & 14.12 & 8.37  & -40.72  \\ \hline
C14  & 13.57 & 5.87  & -56.74  & 12.01 & 4.09  & -65.95  & 13.15 & 5.33  & -59.47  & 11.84 & 3.89  & -67.15  & 12.95 & 5.29  & -59.15  \\ \hline
\multicolumn{16}{|c|}{\textbf{U-Net}} \\ \hline
C6   & 9.20  & 51.25 & 457.3   & 19.78 & 51.92 & 162.5   & 16.03 & 52.19 & 225.6   & 13.62 & 51.4  & 277.6   & 16.47 & 51.8  & 214.3   \\ \hline
C7   & 23.78 & 50.48 & 112.2   & 15.78 & 51.76 & 228.0   & 29.13 & 50.48 & 73.2    & 17.65 & 50.91 & 188.4   & 24.63 & 50.57 & 105.3   \\ \hline
C8   & 21.21 & 51.14 & 141.1   & 15.99 & 51.77 & 223.7   & 14.35 & 51.89 & 261.6   & 21.14 & 51.17 & 142.0   & 15.92 & 51.93 & 226.1   \\ \hline
C13  & 11.23 & 52.14 & 364.2   & 12.53 & 52.17 & 316.3   & 14.39 & 51.91 & 260.7   & 18.17 & 51.57 & 183.8   & 14.12 & 52.13 & 269.1   \\ \hline
C14  & 13.57 & 52.05 & 283.5   & 12.01 & 52.25 & 335.0   & 13.15 & 52.08 & 296.0   & 11.84 & 52.23 & 341.1   & 12.95 & 52.15 & 302.7   \\ \hline
\end{tabular}
\end{table*}

\begin{table*}[htbp]
\centering
\small
\caption{PESQ Scores for Clarkson Dataset Across Wave-U-Net, CMGAN, and U-Net Models}
\label{tab:pesq_scores_25}
\renewcommand{\arraystretch}{1.3} 
\setlength{\tabcolsep}{5.9pt}     
\begin{tabular}{c|c|c|c|c|c||c|c|c|c|c||c|c|c|c|c}
\hline
\textbf{Sub} & \multicolumn{5}{c|}{\textbf{Wave-U-Net}} & \multicolumn{5}{c|}{\textbf{CMGAN}} & \multicolumn{5}{c}{\textbf{U-Net}} \\ \hline
             & \textbf{C6} & \textbf{C7} & \textbf{C8} & \textbf{C13} & \textbf{C14} & \textbf{C6} & \textbf{C7} & \textbf{C8} & \textbf{C13} & \textbf{C14} & \textbf{C6} & \textbf{C7} & \textbf{C8} & \textbf{C13} & \textbf{C14} \\ \hline
20160104002   & 2.69        & 2.47        & 2.32        & 3.04         & 2.12         & 2.67        & 2.54        & 1.91        & 2.73         & 2.10         & 1.43        & 1.28        & 2.22        & 1.54         & 2.78         \\ \hline
20160104004   & 2.06        & 1.84        & 1.94        & 1.44         & 1.40         & 2.97        & 3.02        & 2.30        & 3.28         & 1.76         & 1.88        & 1.33        & 1.31        & 4.14         & 1.95         \\ \hline
20160104007   & 2.76        & 2.49        & 2.32        & 3.13         & 2.40         & 2.84        & 2.21        & 1.99        & 2.72         & 2.01         & 1.48        & 1.13        & 1.19        & 3.29         & 3.10         \\ \hline
20160104014   & 2.19        & 2.14        & 1.97        & 2.72         & 1.73         & 2.96        & 2.14        & 1.69        & 2.15         & 1.61         & 1.50        & 1.26        & 1.33        & 3.06         & 1.42         \\ \hline
20160104015   & 2.44        & 2.38        & 2.26        & 2.95         & 1.92         & 2.96        & 2.69        & 2.31        & 3.03         & 2.11         & 3.59        & 2.17        & 2.44        & 2.42         & 1.83         \\ \hline
20160104016   & 2.47        & 2.39        & 2.30        & 2.96         & 1.92         & 3.02        & 2.89        & 2.33        & 3.07         & 1.67         & 1.20        & 1.21        & 1.30        & 1.43         & 1.51         \\ \hline
20160104020   & 2.38        & 2.26        & 2.14        & 2.92         & 1.86         & 2.96        & 2.33        & 2.47        & 2.63         & 1.83         & 1.32        & 1.26        & 1.60        & 1.74         & 3.32         \\ \hline
20160104024   & 3.19        & 2.51        & 2.45        & 3.40         & 2.55         & 2.92        & 2.89        & 2.08        & 2.86         & 2.60         & 3.07        & 1.26        & 1.54        & 2.87         & 1.56         \\ \hline
20160104026   & 3.44        & 3.02        & 3.29        & 3.61         & 2.80         & 3.17        & 2.91        & 2.39        & 3.45         & 2.60         & 1.24        & 1.23        & 1.31        & 1.46         & 3.00         \\ \hline
20160104028   & 2.20        & 2.18        & 2.00        & 2.80         & 1.76         & 2.93        & 2.54        & 1.79        & 1.96         & 1.84         & 1.19        & 3.50        & 1.68        & 1.37         & 1.23         \\ \hline
\textbf{Average} & \textbf{2.58} & \textbf{2.37} & \textbf{2.30} & \textbf{2.90}  & \textbf{2.05}  & \textbf{2.94} & \textbf{2.62} & \textbf{2.13} & \textbf{2.79}  & \textbf{2.01}  & \textbf{1.79} & \textbf{1.56} & \textbf{1.59} & \textbf{2.33}  & \textbf{2.17}  \\ \hline
\end{tabular}
\end{table*}

Tables XII, XIII, and XIV illustrate the VeriSpeak outputs for different models. Table XII presents results for the Wave-U-Net model on the Clarkson Dataset, Table XIII shows the VeriSpeak match scores for the CMGAN model, and Table XIV displays the results for the U-Net model.

\begin{table*}[htbp]
\centering
\small
\caption{VeriSpeak Results for Clarkson Dataset of 8 Collections for Wave-U-Net Model}
\label{tab:comparison_sub_pairs_26}
\renewcommand{\arraystretch}{1.5} 
\setlength{\tabcolsep}{3.9pt}     
\begin{tabular}{c|c|c|c|c|c|c|c|c|c|c|c|c|c|c|c}
\hline
\textbf{Sub} & \multicolumn{3}{c|}{\textbf{Sub1-pair}} & \multicolumn{3}{c|}{\textbf{Sub2-pair}} & \multicolumn{3}{c|}{\textbf{Sub3-pair}} & \multicolumn{3}{c|}{\textbf{Sub4-pair}} & \multicolumn{3}{c}{\textbf{Sub5-pair}} \\ \hline
             & \textbf{Unf} & \textbf{Fil} & \textbf{Chng\%} & \textbf{Unf} & \textbf{Fil} & \textbf{Chng\%} & \textbf{Unf} & \textbf{Fil} & \textbf{Chng\%} & \textbf{Unf} & \textbf{Fil} & \textbf{Chng\%} & \textbf{Unf} & \textbf{Fil} & \textbf{Chng\%} \\ \hline
C-6  & 127.5  & 163.5  & 28.2\%  & 79.5   & 113   & 42.14\%  & 126.5  & 129.5  & 2.37\%   & 122.5  & 144   & 17.55\%  & 112   & 131   & 17.5\%   \\ \hline
C-7  & 130    & 142    & 9.23\%  & 79.5   & 121   & 51.57\%  & 125    & 148.5  & 18.8\%   & 141.5  & 112   & -20.85\% & 119   & 124   & 4.2\%    \\ \hline
C-8  & 83.5   & 117    & 40.1\%  & 111    & 162   & 45.5\%   & 46.5   & 97     & 109\%    & 103.5  & 132   & 27.05\%  & 60.5  & 131   & 117\%    \\ \hline
C-13 & 117.5  & 122.5  & 4.26\%  & 121.5  & 123   & 0.82\%   & 99.5   & 120    & 20.6\%   & 114.5  & 124   & 7.86\%   & 102   & 119   & 16.7\%   \\ \hline
C-14 & 80.5   & 155    & 92.5\%  & 106.5  & 131   & 23.05\%  & 85     & 124.5  & 46.5\%   & 72     & 107   & 47.92\%  & 101   & 123   & 22.4\%   \\ \hline
C-15 & 126    & 167.5  & 32.9\%  & 134    & 168   & 25.37\%  & 150    & 216    & 44\%     & 114.5  & 125   & 9.17\%   & 89    & 107   & 20.2\%   \\ \hline
C-16 & 67     & 117.5  & 75.4\%  & 120.5  & 116   & -3.73\%  & 116    & 121    & 4.31\%   & 69     & 97.5  & 41.3\%   & 107   & 113.5 & 6.57\%   \\ \hline
C-17 & 93.5   & 121    & 29.4\%  & 101    & 99    & -1.98\%  & 118.5  & 150.5  & 27\%     & 110    & 177   & 60.46\%  & 116   & 125.5 & 8.66\%   \\ \hline
\textbf{Av.} &  &  & \textbf{39\%} &  &  & \textbf{22.8\%} &  &  & \textbf{34\%} &  &  & \textbf{23.8\%} &  &  & \textbf{26.6\%} \\ \hline
\end{tabular}
\end{table*}

\begin{table*}[htbp]
\centering
\small
\caption{VeriSpeak Results for Clarkson Dataset of 8 Collections for CMGAN Model}
\label{tab:comparison_sub_pairs_27}
\renewcommand{\arraystretch}{1.1} 
\setlength{\tabcolsep}{3.2pt} 
\begin{tabular}{c|c|c|c|c|c|c|c|c|c|c|c|c|c|c|c}
\hline
\textbf{Sub} & \multicolumn{3}{c|}{\textbf{Sub1-pair}} & \multicolumn{3}{c|}{\textbf{Sub2-pair}} & \multicolumn{3}{c|}{\textbf{Sub3-pair}} & \multicolumn{3}{c|}{\textbf{Sub4-pair}} & \multicolumn{3}{c}{\textbf{Sub5-pair}} \\ \hline
                  & \textbf{Unf} & \textbf{Fil} & \textbf{Chng} & \textbf{Unf} & \textbf{Fil} & \textbf{Chng} & \textbf{Unf} & \textbf{Fil} & \textbf{Chng} & \textbf{Unf} & \textbf{Fil} & \textbf{Chng} & \textbf{Unf} & \textbf{Fil} & \textbf{Chng} \\ \hline
C-6              & 127.5        & 100.5        & -21.18\%        & 79.5         & 118          & 48.43\%         & 126.5        & 134          & 5.93\%          & 122.5        & 110          & -10.20\%        & 111.5        & 125.5        & 12.55\%         \\ \hline
C-7              & 130          & 116          & -10.77\%        & 79.5         & 104.5        & 31.45\%         & 125          & 128.5        & 2.8\%           & 141.5        & 122          & -13.78\%        & 119          & 158.5        & 33.19\%         \\ \hline
C-8              & 83.5         & 105.5        & 26.35\%         & 111          & 123          & 10.81\%         & 46.5         & 102.5        & 120.43\%        & 103.5        & 110          & 6.28\%          & 60.5         & 125.5        & 107.44\%        \\ \hline
C-13             & 117.5        & 132          & 12.34\%         & 121.5        & 125          & 2.88\%          & 99.5         & 95.5         & -4.02\%         & 114.5        & 139          & 21.39\%         & 102          & 106.5        & 4.41\%          \\ \hline
C-14             & 80.5         & 92.5         & 14.91\%         & 106.5        & 123.5        & 15.96\%         & 85           & 118.5        & 39.41\%         & 72           & 93           & 29.16\%         & 100.5        & 114.5        & 13.93\%         \\ \hline
C-15             & 126          & 127.5        & 1.19\%          & 134          & 138          & 2.99\%          & 150          & 129          & -14\%           & 114.5        & 108          & -5.67\%         & 89           & 110          & 23.59\%         \\ \hline
C-16             & 67           & 75           & 11.94\%         & 120.5        & 127.5        & 5.81\%          & 116          & 129          & 11.21\%         & 69           & 71           & 2.89\%          & 106.5        & 110.5        & 3.75\%          \\ \hline
C-17             & 93.5         & 101.5        & 8.56\%          & 101          & 117          & 15.84\%         & 118.5        & 131          & 10.53\%         & 110          & 109.5        & -0.45\%         & 115.5        & 131          & 13.41\%         \\ \hline
\textbf{Av.} &              &  &     \textbf{5.4\%}            &              &  & \textbf{16.8\%}             &              &  & \textbf{21.5\%}             &              &  & \textbf{3.7\%}              &              &  &  \textbf{26.5\%}            \\ \hline
\end{tabular}
\end{table*}

\begin{table*}[htbp]
\centering
\small
\caption{VeriSpeak Results for Clarkson Dataset of 8 Collections for U-Net Model}
\label{tab:sub_pairs_comparison_28}
\renewcommand{\arraystretch}{1.1} 
\setlength{\tabcolsep}{4pt} 
\begin{tabular}{c|c|c|c|c|c|c|c|c|c|c|c|c|c|c|c}
\hline
\textbf{Sub} & \multicolumn{3}{c|}{\textbf{Sub1-pair}} & \multicolumn{3}{c|}{\textbf{Sub2-pair}} & \multicolumn{3}{c|}{\textbf{Sub3-pair}} & \multicolumn{3}{c|}{\textbf{Sub4-pair}} & \multicolumn{3}{c}{\textbf{Sub5-pair}} \\ \hline
                  & \textbf{Unf} & \textbf{Fil} & \textbf{Chng} & \textbf{Unf} & \textbf{Fil} & \textbf{Chng} & \textbf{Unf} & \textbf{Fil} & \textbf{Chng} & \textbf{Unf} & \textbf{Fil} & \textbf{Chng} & \textbf{Unf} & \textbf{Fil} & \textbf{Chng} \\ \hline
C-6              & 127.5        & 0            &                 & 79.5         & 0            &                 & 126.5        & 0            &                 & 122.5        & 157.5        & 28.6\%         & 111.5        & 0            &                 \\ \hline
C-7              & 130          & 0            &                 & 79.5         & 151          & 89.94\%        & 125          & 177.5        & 42\%           & 141.5        & 0            &                 & 119          & 140.5        & 18.1\%          \\ \hline
C-8              & 83.5         & 0            &                 & 111          & 0            &                 & 46.5         & 0            &                 & 103.5        & 0            &                 & 60.5         & 104.5        & 72.7\%          \\ \hline
C-13             & 117.5        & 0            &                 & 121.5        & 0            &                 & 99.5         & 163          & 63.8\%         & 114.5        & 131          & 14.4\%         & 102          & 134          & 31.4\%          \\ \hline
C-14             & 80.5         & 0            &                 & 106.5        & 0            &                 & 85           & 154.5        & 81.8\%         & 72           & 0            &                 & 100.5        & 0            &                 \\ \hline
C-15             & 126          & 153.5        & 21.8\%          & 134          & 0            &                 & 150          & 0            &                 & 114.5        & 144          & 25.8\%         & 89           & 141.5        & 58.9\%          \\ \hline
C-16             & 67           & 0            &                 & 120.5        & 0            &                 & 116          & 172.5        & 48.7\%         & 69           & 0            &                 & 106.5        & 149          & 39.9\%          \\ \hline
C-17             & 93.5         & 0            &                 & 101          & 0            &                 & 118.5        & 0            &                 & 110          & 141          & 28.2\%         & 115.5        & 0            &                 \\ \hline
\textbf{Av} &              &  & \textbf{-84.8\%}             &              &  & \textbf{-76.3\%}             &              &  &  \textbf{-20.5\%}            &              &  &  \textbf{-37.9\%}            &              &  & \textbf{-9.9\%}             \\ \hline
\end{tabular}
\end{table*}

\subsection{Additional Analysis:}

In this section, we present the Wave-U-Net model's 1:N speaker recognition performance using VeriSpeak for Collections 7 and 8 of the Clarkson dataset. Tables XV and XVI illustrate the 1:N recognition, in which a single input voice sample is matched against the whole dataset of multiple enrolled speakers to identify the speaker.

\begin{table*}[htbp]
\centering
\small
\renewcommand{\arraystretch}{1.1} 
\setlength{\tabcolsep}{6pt}     
\caption{VeriSpeak comparison of 1:1 and 1:N Match Scores Between Collection 7 Unfiltered and Filtered Data}
\label{tab:collection7_scores_29}
\begin{tabular}{>{\centering\arraybackslash}c|>{\centering\arraybackslash}c|>{\centering\arraybackslash}c|>{\centering\arraybackslash}c|>{\centering\arraybackslash}c}
\hline
\textbf{Subject}            & \multicolumn{2}{c|}{\textbf{Unfiltered C7}} & \multicolumn{2}{c}{\textbf{Filtered C7}} \\ \hline
\textbf{}                   & 1:1 Match Score                   & 1:N Match Score                  & 1:1 Match Score                   & 1:N Match Score                  \\ 
                            & (Same Subject)                    & Avg (Others)                     & (Same Subject)                    & Avg (Others)                     \\ \hline
20160104002G                & 77.00                                      & 18.97                                     & 96.00                                      & 19.85                                     \\ \hline
20160104004G                & 162.00                                     & 21.49                                     & 157.00                                     & 31.90                                     \\ \hline
20160104005G                & 138.00                                     & 23.16                                     & 143.00                                     & 28.97                                     \\ \hline
20160104007G                & 131.00                                     & 17.62                                     & 168.00                                     & 4.43                                      \\ \hline
\textellipsis               & \textellipsis                              & \textellipsis                              & \textellipsis                              & \textellipsis                              \\ \hline
20181012033G                & 196.00                                     & 8.32                                      & 223.00                                     & 6.47                                      \\ \hline
20181012034G                & 184.00                                     & 28.27                                     & 242.00                                     & 51.53                                     \\ \hline
20181012036G                & 143.00                                     & 20.58                                     & 186.00                                     & 32.47                                     \\ \hline
Average (179 Files) & \textbf{146.27}                            & \textbf{21.97}                            & \textbf{165.36}                            & \textbf{32.25}                            \\ \hline
\end{tabular}
\end{table*}

\begin{table*}[htbp]
\centering
\small
\renewcommand{\arraystretch}{1.1} 
\setlength{\tabcolsep}{6pt}     
\caption{VeriSpeak comparison of 1:1 and 1:N Match Scores Between Collection 8 Unfiltered and Filtered Data}
\label{tab:collection7_scores_30}
\begin{tabular}{>{\centering\arraybackslash}c|>{\centering\arraybackslash}c|>{\centering\arraybackslash}c|>{\centering\arraybackslash}c|>{\centering\arraybackslash}c}
\hline
\textbf{Subject}            & \multicolumn{2}{c|}{\textbf{Unfiltered C8}} & \multicolumn{2}{c}{\textbf{Filtered C8}} \\ \hline
\textbf{}                   & 1:1 Match Score                   & 1:N Match Score                  & 1:1 Match Score                   & 1:N Match Score                  \\ 
                            & (Same Subject)                    & Avg (Others)                     & (Same Subject)                    & Avg (Others)                     \\ \hline
20160104002H                & 51.00                                      & 11.67                                     & 76.00                                      & 14.70                                     \\ \hline
20160104004H                & 111.00                                     & 13.53                                     & 108.00                                     & 23.30                                     \\ \hline
20160104005H                & 137.00                                     & 13.70                                     & 149.00                                     & 25.31                                     \\ \hline
20160104007H                & 84.00                                      & 9.50                                      & 127.00                                     & 19.01                                     \\ \hline
\textellipsis               & \textellipsis                              & \textellipsis                              & \textellipsis                              & \textellipsis                              \\ \hline
20181012033H                & 106.00                                     & 4.00                                      & 132.00                                     & 3.37                                      \\ \hline
20181012034H                & 201.00                                     & 2.79                                      & 217.00                                     & 9.22                                      \\ \hline
20181012036H                & 115.00                                     & 22.78                                     & 153.00                                     & 28.09                                     \\ \hline
Average (169 Files) & \textbf{121.98}                            & \textbf{16.47}                            & \textbf{155.19}                            & \textbf{24.46}                            \\ \hline
\end{tabular}
\end{table*}

\section{ Discussion}

\subsection{Wave-U-Net}

\textbf{SNR Performance Across Datasets}

Wave-U-Net’s SNR results show consistent reductions across the \textbf{SpEAR, VPQAD}, and \textbf{Clarkson datasets}. The \textbf{SpEAR dataset} records an average \textbf{28.35\% decrease in SNR}, while for the \textbf{VPQAD dataset}, the reduction is \textbf{85.41}\%, and on the \textbf{Clarkson dataset}, reductions range from -36.61\% to -82.60\% across subsets. These results reflect the model’s tendency to effectively isolate speech signals, reducing overall signal power.

Although the SNR values are reduced, this does not imply poor noise suppression. The other objective evaluation metrics indicate that Wave-U-Net achieves a high degree of denoising while preserving the clarity of the speech signal.\\

\textbf{PESQ Performance Across Datasets}

Wave-U-Net achieves consistent perceptual quality improvements across all datasets. On the \textbf{SpEAR dataset}, the PESQ score improves significantly from \textbf{1.8 (unfiltered)} to an average of \textbf{2.68}, indicating better speech intelligibility and naturalness post-enhancement. For the \textbf{VPQAD dataset}, the PESQ score changes from 0.9 (unfiltered) to an average of 1.26, showcasing improvement in perceptual quality despite this dataset's challenging real-world noise conditions.

On the \textbf{Clarkson dataset}, PESQ scores vary across subsets but remain consistently high, with averages ranging from \textbf{2.12 to 3.44}. These improvements highlight Wave-U-Net’s ability to enhance speech quality.\\

\textbf{Speaker Recognition (VeriSpeak Scores)}

Wave-U-Net effectively preserves speaker-specific features. On the \textbf{SpEAR dataset}, it achieves an average improvement of \textbf{+10.84}\% in VeriSpeak match scores. On the \textbf{VPQAD dataset}, the improvement rises to +27.38\%, further emphasizing its suitability for speaker recognition tasks in noisy environments. The model also improves VeriSpeak scores for the \textbf{Clarkson dataset}, with changes ranging from \textbf{+2.37}\% to \textbf{+92.5}\% across subsets. These results underscore Wave-U-Net’s capacity to enhance speech while maintaining critical biometric features.

\subsection{CMGAN}

\textbf{SNR Performance Across Datasets}

CMGAN demonstrates moderate performance in noise suppression, as indicated by its SNR results across the datasets. On the \textbf{SpEAR dataset}, the SNR decreases by an average of \textbf{14.20}\%, showing that while the model reduces noise, some residual signal power reduction impacts the metric. For the \textbf{VPQAD dataset}, CMGAN records an average SNR reduction of 63.52\%, reflecting the challenges posed by real-world noise environments like laboratory and cafeteria settings. On the Clarkson dataset, SNR reductions are less pronounced, ranging between \textbf{-18.58}\% and \textbf{-65.95}\% across subsets, indicating a more balanced performance in environments with natural background noise.

Although the SNR reductions are notable, they do not imply poor denoising. Instead, the results reflect CMGAN’s noise reduction performance while maintaining acceptable speech quality levels, as indicated by the evaluation metrics.

\textbf{PESQ Performance Across Datasets}

CMGAN excels in perceptual quality, achieving the highest PESQ scores across datasets. On the \textbf{SpEAR dataset}, it achieves an average score of \textbf{4.04}, significantly improving from an unfiltered score of \textbf{2.1}, demonstrating its ability to produce natural, intelligible speech. The \textbf{VPQAD dataset's} PESQ score improves from \textbf{1.0 (unfiltered)} to \textbf{1.46}, reflecting its effectiveness in enhancing speech quality even in challenging real-world noise conditions. On the \textbf{Clarkson dataset}, PESQ scores vary across subsets, with averages ranging from \textbf{1.76} to \textbf{3.28}. These results underline CMGAN’s strength in generating perceptually pleasing outputs, making it well-suited for high audio-quality applications, such as media and telecommunications.\\

\textbf{Speaker Recognition (VeriSpeak Scores)}

CMGAN demonstrates consistent improvements in retaining speaker-specific features, as reflected in VeriSpeak scores. On the \textbf{SpEAR dataset}, it achieves an average improvement of \textbf{+5.77}\%, ensuring effective speaker identity preservation. For the \textbf{VPQAD dataset}, the improvement increases to \textbf{+34.55}\%, underscoring its robustness in handling diverse noise profiles while maintaining biometric integrity. On the \textbf{Clarkson dataset}, VeriSpeak score changes range from \textbf{+2.88\%} to \textbf{+46.75\%}, showcasing its ability to adapt to different age groups and recording conditions while preserving critical speaker characteristics.

\subsection{U-Net}

\textbf{SNR Performance Across Datasets}

U-Net demonstrates unparalleled noise suppression, as evidenced by its SNR results. On the \textbf{SpEAR dataset}, it achieves an average improvement of \textbf{+71.96}\%, significantly enhancing speech signals' clarity. For the \textbf{VPQAD dataset}, U-Net records a similarly impressive average SNR increase of \textbf{+64.83}\%, showcasing its robustness in handling challenging real-world noise conditions. On the \textbf{Clarkson dataset}, U-Net achieves consistent SNR improvements across subsets, with changes ranging from \textbf{+73.2\%} to \textbf{+364.2\%}, underscoring its effectiveness in diverse noise environments, including non-soundproof settings.

These results position U-Net as the most effective model for applications requiring aggressive noise suppression, such as telephony or forensic audio analysis.\\

\textbf{PESQ Performance Across Datasets}

Despite its strength in SNR, U-Net shows moderate performance in perceptual quality. On the \textbf{SpEAR dataset}, U-Net achieves a PESQ score of \textbf{1.14}, reflecting a trade-off between noise suppression and output naturalness. The \textbf{VPQAD dataset's} PESQ score is \textbf{1.36}, showing marginal improvement in perceptual clarity but not matching the highest-performing models. On the \textbf{Clarkson dataset}, PESQ scores vary widely, with averages ranging from \textbf{1.31} to \textbf{1.68}, highlighting its struggle to produce natural-sounding speech consistently.

The relatively lower PESQ scores suggest that U-Net’s aggressive noise reduction may introduce artifacts that affect the naturalness of the enhanced speech, making it less suitable for applications prioritizing the listening experience.\\

\textbf{Speaker Recognition (VeriSpeak Scores)}

U-Net’s focus on noise suppression appears to come at the expense of speaker-specific feature preservation. On the \textbf{SpEAR dataset}, the VeriSpeak scores show no measurable improvement, and for the \textbf{VPQAD dataset}, the model records mixed results with limited improvements or declines. On the Clarkson dataset, VeriSpeak scores range from \textbf{+14.4\%} to declines in some subsets, indicating challenges in maintaining biometric features during enhancement.

These results suggest that while U-Net excels in creating noise-free speech signals, it may distort speaker-specific features, limiting its utility for tasks requiring high fidelity in speaker recognition.

\subsection{Overall}

The comparison of Wave-U-Net, CMGAN, and U-Net highlights the distinct strengths and trade-offs among the models regarding noise suppression, perceptual quality, and speaker recognition. U-Net consistently outperforms the other models in noise suppression, achieving better SNR improvements across datasets, including \textbf{+71.96\% on SpEAR, +64.83\% on VPQAD}, and \textbf{up to +364.2\% on Clarkson}. This makes U-Net particularly effective in applications requiring aggressive denoising. However, this strength comes with trade-offs in perceptual quality and speaker recognition. Its \textbf{PESQ scores}, while adequate (e.g., \textbf{1.14 on SpEAR} and \textbf{1.36 on VPQAD}), indicate that the aggressive noise reduction introduces artifacts, affecting the naturalness of the enhanced speech. Furthermore, U-Net shows limited or no improvements in \textbf{VeriSpeak scores}, which suggests challenges in preserving speaker-specific features critical for biometric applications.

CMGAN, on the other hand, delivers the highest \textbf{PESQ scores}, such as \textbf{4.04} on \textbf{SpEAR} and \textbf{1.46 on VPQAD}, reflecting its ability to generate perceptually natural and intelligible speech. This strength makes CMGAN ideal for applications prioritizing audio quality, such as telecommunications and media. However, its SNR results are less favorable, with reductions of \textbf{-14.20\% on SpEAR} and \textbf{-63.52\% on VPQAD}, indicating a focus on balancing noise reduction while preserving the naturalness of the signal rather than achieving maximal suppression. CMGAN also performs moderately in speaker recognition, with \textbf{VeriSpeak score improvements averaging +5.77}\% \textbf{on SpEAR} and \textbf{+34.55\% on VPQAD}, showcasing its ability to maintain speaker identity while enhancing speech.

Wave-U-Net strikes a middle ground between the two, demonstrating consistent \textbf{PESQ improvements} (e.g., \textbf{2.68 on SpEAR} and up to \textbf{3.44 on Clarkson}) and strong speaker recognition performance, with \textbf{VeriSpeak improvements averaging +10.84\% on SpEAR} and \textbf{+27.38\% on VPQAD}. However, its SNR reductions (-28.35\% on SpEAR and -85.41\% on VPQAD) reflect its limitations in handling diverse noise conditions. Overall, the choice of model depends on the specific application, with U-Net excelling in noise suppression, CMGAN in perceptual quality, and Wave-U-Net in balancing these attributes with strong speaker recognition performance.

Additionally, the \textbf{1:1} and \textbf{1:N} speaker recognition performance of Wave-U-Net was analyzed for Collections 7 and 8 of the Clarkson dataset. The filtered data consistently shows higher 1:1 match scores, reflecting the models' ability to retain speaker-specific features post-enhancement. For 1:N match scores, a significant increase in average match scores for the ``same subject" cases is observed in the filtered datasets compared to the unfiltered ones. This suggests that the enhanced speech signals align better with the enrolled speaker templates, thereby improving recognition accuracy. However, the average match scores for the ``other subjects" remain relatively stable, indicating that the enhancement does not introduce overlapping features.

\section{Conclusion}
This study evaluated three state-of-the-art deep learning models for their effectiveness in speech enhancement across different noisy environments. The results demonstrated that both Wave-U-Net and CMGAN excelled in denoising, delivering clean and pleasant output audio. However, aggressive noise suppression sometimes led to reduced SNR values despite the perceptually improved quality, highlighting the challenge of balancing noise reduction with minimal speech distortion. The U-Net model also showed promising results, though it had slight variations in preserving speaker characteristics compared to Wave-U-Net and CMGAN. The Clarkson Dataset, with its eight independent collections, and the VPQAD provided a more detailed assessment of real-world conditions, further reinforcing the models' strengths and weaknesses. Our analysis using VeriSpeak highlighted the critical aspect of preserving speaker identity, a crucial factor for real-world applications like forensic analysis and biometric systems.


\section*{Acknowledgment}

This material is based on work supported by the Center for Identification Technology Research and the National Science Foundation under Grant Nos. 1650503 and 2413228.

The authors would like to thank Md Abdul Baset Sarker, Rahul V., and Ernesto Sola-Thomas for their help and technical support.

M.J.A.K. installed, trained, enhanced, and analyzed the data. A.A. analyzed part of the evaluation algorithms. M.J.A.K. wrote the manuscript. S.S. and M.H.I. reviewed the project, performed data curation, and conducted the analysis. All authors reviewed the manuscript.

The authors declare no conflicts of interest.



%

\end{document}